\documentclass[final,3p,times]{elsarticle}
\pdfoutput=1
\usepackage{bbding}
\usepackage{pifont}
\usepackage{amsfonts}
\usepackage{mathrsfs}
\usepackage{amsmath}
\usepackage[amsmath]{ntheorem}
\usepackage{graphicx}
\usepackage{subfig}
\usepackage{epstopdf}
\usepackage{mathtools}
\usepackage{color}
\usepackage{arydshln}
\usepackage{amssymb}
\usepackage{booktabs}
\usepackage{bookmark}
\usepackage{tikz}
\usepackage{multirow}
\usepackage{appendix}
\usepackage{xpatch} %color

\newcommand*{\circled}[1]{\lower.7ex\hbox{\tikz\draw (0pt, 0pt)%
		circle (.5em) node {\makebox[1em][c]{\small #1}};}} %圆圈数字

\newtheorem{theorem}{Theorem}

\newtheorem{assumption}{Assumption}
\newtheorem{remark}{Remark}
\newtheorem{definition}{Definition}
\newproof{Proof}{\bf{Proof}}

\biboptions{sort&compress}

\journal{ }

\begin{document}
	
	\begin{frontmatter}
		
		
\author[1]{Jiamin Wang}
\ead{wangjiamin21@outlook.com}
% \author[2]{Liqi Zhou}
% \ead{lqzhou\_96@foxmail.com}
% \author[3]{Dong Zhang}
% \ead{zhangdong@nwpu.edu.cn}
\author[1]{Jian Liu}
\ead{liujianzym@outlook.com}
% \author[4]{Feng Xiao}
% \ead{fengxiao@ncepu.edu.cn}
\author[2]{Feng Xiao}
\ead{fengxiao@ncepu.edu.cn}
\author[3]{Ning Xi}
\ead{nxi@xidian.edu.cn}
\author[1]{Yuanshi Zheng\corref{cor1}}
\ead{zhengyuanshi2005@163.com}
\cortext[cor1]{Corresponding author}

\affiliation[1]{
	organization={Shaanxi Key Laboratory of Space Solar Power Station System, School of Mechano-electronic Engineering, Xidian University},
	%	addressline={},
	city={Xi'an},
	postcode={710071},
	%	state={shannxi},
	country={China}}
\affiliation[2]{
	organization={School of Control and Computer Engineering, North China Electric Power University},
	%	addressline={},
	city={Beijing},
	postcode={102206},
	%	state={shannxi},
	country={China}}
\affiliation[3]{
	organization={School of Cyber Engineering, Xidian University},
	%	addressline={},
	city={Xi'an},
	postcode={710071},
	%	state={shannxi},
	country={China}}
		% \affiliation[2]{
			% 	organization={School of Astronautics, Northwestern Polytechnical University},
			%	%	addressline={},
			%	city={Xi'an},
			%	postcode={710072},
			%	%	state={shannxi},
			%	country={China}}

		\title{Scalable second-order consensus of hierarchical groups\tnoteref{t1}}
		\tnotetext[t1]{This work is funded by the National Natural Science Foundation of China  (62273267), the Natural Science Basic Research Program of Shaanxi, China (2022JC-46, 2023-JC-YB-525 and 2023-JC-QN-0766), the Fundamental Research Funds for the Central Universities, China (ZYTS23021), and the Major Research plan of the National Natural Science Foundation of China (92267204).}
		
		%\author{}
		%
		%\affiliation{organization={},%Department and Organization
		
		\begin{abstract}
			Motivated by widespread dominance hierarchy, growth of group sizes, and feedback mechanisms in social species,
			we are devoted to exploring the scalable second-order consensus of hierarchical groups.
			More specifically, a hierarchical group consists of a collection of agents with double-integrator dynamics on a directed acyclic graph with additional reverse edges, which characterize feedback mechanisms across hierarchical layers.
			%Scalable second-order consensus refers to maintaining second-order consensus without re-tuning the control gains as the group size grows and the reverse edges emerge.
			As the group size grows and the reverse edges appears,
			we investigate whether the absolute velocity protocol and the relative velocity protocol can preserve the system consensus property without tuning the control gains. 
			%We investigate whether the absolute velocity protocol and the relative velocity protocol can preserve scalable second-order consensus.
			It is rigorously proved that the absolute velocity protocol is able to achieve completely scalable second-order consensus but the relative velocity protocol cannot.
			This result theoretically reveals how the scalable coordination behavior in hierarchical groups is determined by local interaction rules.
			Moreover, we develop a hierarchical structure in order to achieve scalable second-order consensus for networks of any sizes and with any number of reverse edges.
		\end{abstract}
		
		
		
	\end{frontmatter}
	
	
	\section{Introduction}\label{Sec:1}
	The dominance hierarchy is a pyramidal social relationship formed by ranking individuals according to some criteria such as the orders of motion, divisions of labor, body sizes, health conditions, and leadership,
	which is omnipresent in social species such as bird flocks \cite{Nagy2010}, ant colonies \cite{Franks1983}, and primate groups \cite{Sapolsky2005} including human society.
	A mathematical model for characterizing such hierarchical relationships is directed acyclic graphs (DAGs) \cite{Nagy2010,Shimoji2014,Gross2003},
	in which all edges originate from lower-numbered to higher-numbered vertices and no directed cycles are included.
	It indicates that higher-ranked individuals exert greater authority to lead lower-ranked individuals.
	Interestingly, the hierarchy facilitates highly efficient coordination behavior.
	For example,
	in hierarchical pigeon flocks \cite{Nagy2010},
	a rapid consensus of the flight directions emerged.
	Thus, it has attracted substantial interest from researchers in the filed of the distributed coordination of multi-agent systems.
	Based on multi-agent models, authors in \cite{Shao2016,Shao2018} theoretically proved that hierarchical networks described by DAGs indeed lead to optimal convergence rates of consensus.

	Interestingly, we notice that feedback exists from low-ranked layers to high-ranked layers in hierarchical groups.
	Since low-ranked individuals usually undertake foraging and exploration tasks,
	they should report feedback information such as food resources, suitable habitats and dangerous predators to high-ranked individuals for better collective behavior decisions \cite{Detrain2006}.
	Moreover, the hierarchy of most gregarious animals is not strictly linear \cite{Shimoji2014,Douglas2017},
	that is,
	the hierarchical relationship between a high-ranked individual and a low-ranked individual is temporarily reverse sometimes.
	Such a feedback mechanism, which is opposite to the hierarchical relationship,
	can be viewed as additional directed edges in DAGs.
	These edges point from higher-numbered to lower-numbered vertices.
	Thereby, they are called the \textit{reverse edges} that may create directed cycles and affect the consensus performance.
	Inspired by these phenomena, some researchers examined the influence of adding reverse edges to DAGs on the convergence rates of first-order consensus \cite{Zhang2017,Mo2019,Hao2019,Zhang2022}.
	It should be noted that first-order consensus is always achieved no matter how many reverse edges are added to a DAG since the underlying interaction graphs always contain spanning trees.
	By contrast, 
	second-order consensus is sensitive to the complex eigenvalues of the Laplacian matrix \cite{Zhu2009,Yu2010}.
	Additional reverse edges may bring complex eigenvalues and thus second-order consensus could fail without adjusting control gains.
	In order to recapture consensus, authors in \cite{Dubey2022,Dubey2023} suggested re-tuning control gains or modifying edge weights.
	Note that the above results were derived on particular regular DAGs and supposed no more than two reverse edges.
	However, hierarchical networks of large-scale groups are asymmetric,
	and feedback mechanisms emerge quite frequently that is far beyond the assumption of two reverse edges.
	Therefore, it is necessary to investigate the second-order consensus problems of hierarchical groups on general DAGs in the presence of multiple reverse edges.
	
	More interestingly, it is observed that when an incipient hierarchical group grows and matures by reproduction,
	the hierarchical network has to be re-ranked to accommodate the offspring \cite{Fischhoff2007}.
	Furthermore, feedback mechanisms emerge and vary more frequently as the group grows and the environment changes \cite{Shimoji2014,Franks1983}.
	In this case, 
	in order to rebuild consensus under constantly changing network structure together with reverse edges,
	we have to re-tune control gains or edge weights as shown in \cite{Dubey2022,Dubey2023},
	which is not practical for a group with finite network resources.
	On the other hand, a recent work \cite{Tegling2023} focused on the growth of network sizes for multi-agent systems,
	which allowed more agents to join but the control gains are preset and cannot be re-tuned.
	Based on these observations, we study the \textit{scalable second-order consensus} of hierarchical groups, that is, reaching consensus without re-tuning the control gains in the presence of reverse edges as the group size grows.
	
	On the other hand,
	the authors in \cite{Tegling2023} showed that the consensus vanished for certain large-scale networks with local relative information feedback.
	Similarly, in aforementioned works \cite{Dubey2022,Dubey2023},
	the classic relative velocity protocol \cite{Ren2007} was investigated to study the influence of reverse edges on second-order consensus,
	where each agent uses the relative velocity measurements with regard to its neighbors.
	In contrast with the relative velocity protocols,
	another types of second-order consensus protocols,
	known as the absolute velocity protocol \cite{Xie2007},
	required each agent to use their own absolute velocity information as the local feedback.
	Then, based on the structures of above two state-of-the-art protocols,
	a large number of investigations of second-order consensus have been conducted  \cite{Zheng2011,Li2015,Hou2017,Zheng2019,Zhao2020-1,Xu2022,Zhou2023}.
	Recently, the authors in \cite{Wang2022} gave a concise graph condition to tell which one of the absolute and relative velocity protocols has a better anti-disturbance capability.
	Based on these observations, we believe that the underlying structures of protocols may determine the scalable second-order consensus of hierarchical groups.
	It is consistent with the phenomenon that ant colonies would change the local interaction rules for larger scale of coordinated foraging \cite{Detrain2006}.
	Therefore, in this article, we compare the absolute velocity protocol and the relative velocity protocol with respect to the scalable second-order consensus.
	As far as we known, no previous study has investigated this issue.
	%It is consistent with the phenomenon that the scale of coordination behaviors is limited by the local interaction rules.
	
	Motivated by the above works,
	we delve into the scalable second-order consensus of hierarchical groups,
	which is composed of a group of agents with double-integrator dynamics on DAGs.
	We investigate which one of the absolute velocity protocols and the relative velocity protocols can achieve scalable second-order consensus.
	The considered problem is challenging.
	Because the existing methods in aforementioned references \cite{Zhang2017,Mo2019,Hao2019,Zhang2022,Dubey2022,Dubey2023,Tegling2023}
	depended on precise Laplacian eigenvalues or specific topological structures,
	it is difficult to derive the sufficient conditions for scalable second-order consensus on general DAGs with multiple reverse edges.
	The main contributions are in three aspects:
	\begin{itemize}
		\item Different from existing works that studied symmetric DAGs with two or less reverse edges,
		we consider the general DAGs with multiple reverse edges;
		\item We prove that under a certain assumption,
		the absolute velocity protocols can obtain scalable second-order consensus no matter how the hierarchical network and the reverse edges change but the relative velocity protocols cannot.
		This result reveals the pivotal role of underlying structures of protocols on the scalable consensus of hierarchical groups.
		It is beneficial to explain why the local interaction rules for coordination behaviors are changed when a small hierarchical group grows to a large-scale group \cite{Detrain2006};
		\item We propose a hierarchical structure, i.e., directed star graphs,
		on which the scalable second-order consensus can be guaranteed under reverse edges of any number and any weight.
		This finding facilitates deploying large-scale scalable multi-agent systems without re-tuning  the controller parameters and edge weights.
	\end{itemize}
	
	The rest of this article is organized as follows.
	In Section \ref{Sec:2}, we give some preliminaries in graph theory.
	We formulate the problem in Section \ref{Sec:3}.
	The analysis of scalable consensus for the absolute and the relative velocity protocols are shown in Section \ref{Sec:4}.
	%In Section \ref{Sec:4}, we analyze the \Hinf performance of the sampled-data multi-agent network and some extension issues.
	%In Section \ref{Sec:4}, numerical tests are given for several families of communication graphs.
	We conclude our work in Section \ref{Sec:5}.
	%%%%%%%%%%%%%%%%%%%%%%%%%%%%%%%%%%%%%%%%%%%%%%%%%%%%%%%%%%%
	
	\vspace{10pt}
	\noindent\textbf{Notations}:
	Throughout this article, $\mathbb{R}^n$ and $\mathbb{C}^n$ are the
	$n$-dimensional real column vector space and complex column vector space, respectively.
	$\mathbb{R}^{m\times n}$ is the $m\times n$ real matrix 
	space.
	Let $\mathbf{1}$ and $\mathbf{0}$, respectively, 
	be the all-one and all-zero matrices with appropriate dimensions.
	Specially, $\mathbf{1}_n$ and $\mathbf{0}_n$ denote the $n\times1$
	all ones and all zeros column vectors, respectively.
	Denote $I_n$ by the $n$-dimensional identity matrix.
	Define a set $\mathcal{I}_m=\{1,2,\dots,m\}$.
	A complex number $\lambda\in\mathbb{C}$ is denoted by $\lambda=\text{Re}(\lambda)+\mathbf{j}\text{Im}(\lambda)$, where $\text{Re}(\lambda)$ is the real part of $\lambda$, $\text{Im}(\lambda)$ is the imaginary part of $\lambda$, and $\mathbf{j}$ is the imaginary unit.
	$|\cdot|$ refers to the cardinality of a set or the modulus of a complex number.
	%$\mathcal{L}_2\left[0,\infty\right)$ denotes the space of square-integrable signals over $\left[0,\infty\right)$ and $\ell_2\left[0,\infty\right)$ represents the space of square-summable infinite sequences.

	%%%%%%%%%%%%%%%%%%%%%%%%%%%%%%%%%%%%%%%%%%%%%%%%%%%%%%%%%%%%%%%

	%%%%%%%%%%%%%%%%%%%%%%%%%%%%%%%%%%%%%%%%%%%%%%%%%%%%%%%%%%%%%%%
	%%%%%%%%%%%%%%%%%%%%%%%%%%%%%%%%%%%%%%%%%%%%%%%%%%%%%%%%%%%%%%%
	\section{Preliminaries}\label{Sec:2}
	%%%%%%%%%%%%%%%%%%%%%%%%%%%%%%%%%%%%%%%%%%%%%%%%%%%%%%%%%%%%%%%
	%%%%%%%%%%%%%%%%%%%%%%%%%%%%%%%%%%%%%%%%%%%%%%%%%%%%%%%%%%%%%%%
	
	Consider a group of $n>2$ agents whose interaction network is represented as a weighted directed graph $\mathcal{G}_n=(\mathcal{V}_n,\mathcal{E}_n,\mathcal{A}_n)$,
	where $\mathcal{V}_n=\{1,2,\dots,n\}$ is the set of vertices, $\mathcal{E}_n\subseteq\mathcal{V}_n\times\mathcal{V}_n$ is the set of edges, and $\mathcal{A}_n=[a_{ij}]_{n\times n}$ is the weighted adjacency matrix with $a_{ij}\geq 0$.
	An edge of $\mathcal{G}_n$ is denoted by $\varepsilon_{ij}=(i,j)$, where $j$ is called the 
	parent vertex of $i$ and $i$ is the child vertex of $j$.
	$(i,j)\in\mathcal{E}_n$ $\Leftrightarrow$ $a_{ij}>0$ $\Leftrightarrow$ agent $i$ can 
	receive information from agent $j$.
	%The neighbor set of vertex $i$ is indicated by $\mathcal{N}_i=\{j|(i,j)\in\mathcal{E}_n,j\in\mathcal{V}_n\}$.
	Moreover, $\mathcal{G}_n$ has no self-cycles, i.e., $a_{ii}=0$.
	The degree matrix $\mathcal{D}_n=[d_{ij}]_{n\times n}$ is a diagonal matrix with
	$d_{ii}=\sum_{j=1}^na_{ij}$. 
	For convenience, let $d_{\max}=\max_{i\in\mathcal{V}_n}\{d_{ii}\}$.
	The Laplacian matrix $\mathcal{L}_n=[l_{ij}]\in\mathbb{R}^{n\times n}$ associated with the graph $\mathcal{G}_n$ is defined as $l_{ii}=\sum_{j=1, j\neq i}^{n}a_{ij}=d_{ii}$ and $l_{ij}=-a_{ij}$, $j\neq i$, hence $\mathcal{L}_n\mathbf{1}_n=\mathbf{0}_n$ holds.
	Let $\lambda_i$, $i\in\mathcal{V}_n$ be the eigenvalues of $\mathcal{L}_n$.

	The following definitions and assumption will be utilized to establish our main results.
	\begin{definition}[\cite{Gross2003}]\label{def:0}
		A DAG is a directed graph without directed cycles.
	\end{definition}

	A DAG has a \textit{linear extension ordering} which is a consecutive increasing numbering of the vertices as the order $1,2,\dots,n$ such that all directed edges point from lower-numbered to higher-numbered vertices \citep{Gross2003}.
	This property is used to characterize the relationships among members in hierarchical groups.
	Specifically, higher-ranked individuals in the hierarchy correspond to lower-numbered vertices in the linear extension ordering,
	and the directed edges mean that lower-ranked individuals are always led by higher-ranked individuals.
	In addition, the feedback from low-ranked layers to high-ranked layers in hierarchical groups is described by additional reverse edges in DAGs,
	which is defined as follows.
	
	\begin{definition}[\cite{Zhang2017}]\label{def:1}
		For two vertices $i$ and $j$ $(i<j)$ in a DAG $\mathcal{G}_n$ that has a linear extension ordering,
		the additional edge $(i,j)$ is called a \textit{reverse edge} from $j$ to $i$.
	\end{definition}
	%\begin{definition}[\cite{Zhang2022}]\label{def:2}
	%	
	%%	, and $s=l-h$ is called the \textit{reverse range}.
	%\end{definition}
	In other words, a reverse edge is an extra edge that is opposite to the linear extension ordering.
	Generally,
	suppose that there are multiple reverse edges and there is at most one reverse edge between two different vertices.
	Then, the DAG $\mathcal{G}_n$ and additional reverse edges constitute a new directed graph $\bar{\mathcal{G}}_n=(\mathcal{V}_n,\bar{\mathcal{E}}_n,\bar{\mathcal{A}}_n)$ termed the \textit{mixed graph},
	where $\bar{\mathcal{E}}_n=\mathcal{E}_n\bigcup\mathcal{E}^r_n$ is the set of edges,
	$\mathcal{E}_n$ is the set of edges of the original DAG $\mathcal{G}_n$,
	$\mathcal{E}^r_n$ is the set of additional reverse edges,
	and $\bar{\mathcal{A}}_n=[\bar{a}_{ij}]_{n\times n}$ is the weighted adjacency matrix whose elements are defined as $\bar{a}_{ii}=0$ and $\bar{a}_{ij}>0$ if $(i,j)\in\bar{\mathcal{E}}_n$ where $i\neq j$.
	It is obvious that $\bar{a}_{ij}=a_{ij}>0$, $\forall (i,j)\in \mathcal{E}_n$, where $i>j$,
	and $\bar{a}_{ij}>0$, $\forall (i,j)\in \mathcal{E}^r_n$,
	where $i<j$.
	\begin{definition}\label{def:2}
		For a vertex $i$ in a mixed graph $\bar{\mathcal{G}}_n$,
		if there is another vertex $j$ such that $i>j$ and $(i,j)\in \mathcal{E}^r_n$,
		$j$ is called a superior neighbor of $i$.
		If the vertex $j$ satisfies $i<j$ and $(i,j)\in \mathcal{E}^r_n$,
		$j$ is said to be a inferior neighbor of $i$.
	\end{definition}
	Specially, $\mathcal{N}^{h}_i=\{j|(i,j)\in\bar{\mathcal{E}}_n, j\in\mathcal{V}_n, i>j\}$
	and
	$\mathcal{N}^{l}_i=\{j|(i,j)\in\bar{\mathcal{E}}_n, j\in\mathcal{V}_n, i<j\}$
	denote the superior neighbor set and the inferior neighbor set of vertex $i$, respectively.
	% $(h_1,l_1),\dots,(h_m,l_m)$ with positive weights $r_1,\dots,r_m$, where $1\leq h_k<l_k\leq n$, $\forall k\in\mathcal{I}_m$, 
	Let $\theta=\min\{i|(i,j)\in\mathcal{E}^r_n\}$,
	$\varphi=\max\{j|(i,j)\in\mathcal{E}^r_n\}$ and $s=\varphi-\theta$.
	%Furthermore, linear extension ordering means that the Laplacian matrix associated with a DAG can be written as a lower triangular matrix and obviously .
	Then, the Laplacian matrix $\bar{\mathcal{L}}_n=[\bar{l}_{ij}]\in\mathbb{R}^{n\times n}$ of the mixed graph $\bar{\mathcal{G}}_n$ can be written as
	\begin{equation}\label{L_bar}
		\bar{\mathcal{L}}_n=\underbrace{\begin{bmatrix}L_1 & &\\
				* & \Theta & \\
				* & * & L_2\end{bmatrix}}_{\mathcal{L}_n}+
		\underbrace{\begin{bmatrix}\mathbf{0} & &\\
				& \Delta & \\
				&  & \mathbf{0}\end{bmatrix}}_{P},
	\end{equation}
	where 
	$L_1\in\mathbb{R}^{(\theta-1)\times(\theta-1)}$,
	$\Theta\in\mathbb{R}^{(s+1)\times(s+1)}$,
	and $L_2\in\mathbb{R}^{(n-\varphi)\times(n-\varphi)}$
	are all lower triangular matrices,
	$\Delta\in\mathbb{R}^{(s+1)\times(s+1)}$ is an upper triangular matrix,
	$\mathcal{L}_n$ is the Laplacian matrix of the original DAG $\mathcal{G}_n$ with a linear extension ordering,
	and $P=[p_{ij}]\in\mathbb{R}^{n\times n}$ captures the additional reverse edges.
	Specifically, non-diagonal elements of $P$ are given as
	$p_{ij}=-\bar{a}_{ij}$ if $(i,j)\in \mathcal{E}^r_n$,
	otherwise $p_{ij}=0$,
	%there exists a reverse edge $(h_k,l_k)$, $k\in\mathcal{I}_m$ such that $h_k=i$ and $l_k=j$, otherwise $p_{ij}=0$,
	and the diagonal elements of $P$ is $p_{ii}=\sum_{j> i}^{s+1}\bar{a}_{ij}$.
	It is obvious that the eigenvalues of $\mathcal{L}_n$ are $\lambda_i=d_{ii}$, $\forall i\in\mathcal{V}_n$.
	Furthermore, the eigenvalues of $\bar{\mathcal{L}}_n$ are denoted by $\bar{\lambda}_i$, $i\in\mathcal{V}_n$.
	
	The following assumption is made for the mixed graphs throughout this article unless otherwise stated.
	%and $\mathcal{A}^r$ 
	\begin{assumption}\label{assumption:2}
		For any mixed graph $\bar{\mathcal{G}}_n$,
		each agent has finite superior neighbors and inferior neighbors, i.e., $|\mathcal{N}^h_i|\leq\zeta$, $|\mathcal{N}^l_i|\leq\xi$, $\forall i\in\mathcal{V}_n$,
		and all edge weights are finite, i.e.,
		$\bar{a}_{ij}\leq \bar{a}$, $\forall (i,j)\in\mathcal{E}_n$ and 
		$\bar{a}_{ij}\leq \bar{a}_r$, $\forall (i,j)\in\mathcal{E}^r_n$.
	\end{assumption}
	\begin{remark}
		Assumption \ref{assumption:2} is reasonable in large-scale natural groups.
		For example, in fish schools and bird flocks,
		due to the limited spatial communication distance,
		each individual adjusts its direction based on the directions of its finite nearest neighbor,
		and the interaction strengths among individuals are finite and decay with the spatial distance \cite{Vicsek1995,Mora2016}.
	\end{remark}

	%%%%%%%%%%%%%%%%%%%%%%%%%%%%%%%%%%%%%%%%%%%%%%%%%%%%%%%%%%%%%%%%%%%%%%
	%%%%%%%%%%%%%%%%%%%%%%%%%%%%%%%%%%%%%%%%%%%%%%%%%%%%%%%%%%%%%%%%%%%%%%
	\section{Problem statement}\label{Sec:3}
	%%%%%%%%%%%%%%%%%%%%%%%%%%%%%%%%%%%%%%%%%%%%%%%%%%%%%%%%%%%%%%%%%%%%%%
	%%%%%%%%%%%%%%%%%%%%%%%%%%%%%%%%%%%%%%%%%%%%%%%%%%%%%%%%%%%%%%%%%%%%%%
	In this article, the hierarchical group consists of $n$ agents on the DAG $\mathcal{G}_n$,
	which take the following double-integrator dynamics
	\begin{equation}\label{eq:2.2.1}
		\begin{aligned}
			\dot{x}_i(t) & =v_i(t), \\
			\dot{v}_i(t) & =u_i(t), i\in\mathcal{V}_n,
		\end{aligned}
	\end{equation}
	where $x_i(t)\in\mathbb{R}$, $v_i(t)\in\mathbb{R}$, and $u_i(t)\in\mathbb{R}$ are the position-like, the velocity-like, and the control input of agent $i$, respectively,
	and $\mathcal{G}_n$ has a spanning tree and a linear extension ordering.
	The hierarchical group \eqref{eq:2.2.1} is said to reach second-order consensus if and only if $\lim_{t\to \infty}\Vert x_i(t)-x_j(t)\Vert=0$ and $\lim_{t\to \infty}\Vert v_i(t)-v_j(t)\Vert=0$, $\forall i,j\in\mathcal{V}_n$.
	There are two common protocols for second-order consensus,
	the absolute velocity protocol
	\begin{equation}\label{ab_pro}
		u_i(t)\!=\!\alpha\sum_{j=1}^n{a_{ij}\big[x_j(t)-x_i(t)\big]}-\beta v_i(t),
	\end{equation}
	and the relative velocity protocol
	\begin{equation}\label{re_pro}
		u_i(t)\!=\!\alpha\!\sum_{j=1}^n{\! a_{ij}\big[x_j(t)\!-\!x_i(t)\big]}\!+\!\beta \!\sum_{j=1}^n{\!a_{ij}\big[v_j(t)\!-\! v_i(t)\big]},
	\end{equation}
	where $a_{ij}$ is the $(i,j)_{\text{th}}$ entry of the weighted adjacency matrix $\mathcal{A}_n$ associated with the DAG $\mathcal{G}_n$, and the positive constants $\alpha$ and $\beta$ are the state control gain and the velocity control gain, respectively.
	For both protocols, the necessary and sufficient conditions of reaching second-order consensus on directed graphs were investigated in \cite{Zhu2009,Yu2010}.
	Specifically,
	hierarchical group \eqref{eq:2.2.1}-\eqref{ab_pro} reaches second-order consensus if and only if $\mathcal{G}_n$ has a spanning tree and
	\begin{equation}\label{eq:2.2.4}
		\frac{\beta^2}{\alpha}>\max_{\lambda_i\neq 0}\frac{\text{Im}^2(\lambda_i)}{\text{Re}(\lambda_i)}.
	\end{equation}
	Similarly, hierarchical group \eqref{eq:2.2.1}-\eqref{re_pro} reaches second-order consensus if and only if $\mathcal{G}_n$ has a spanning tree and
	\begin{equation}\label{eq:2.2.5}
		\frac{\beta^2}{\alpha}>\max_{\lambda_i\neq 0}\frac{\text{Im}^2(\lambda_i)}{\text{Re}(\lambda_i)|\lambda_i|^2}.
	\end{equation}
	Since $\mathcal{G}_n$ is a DAG, we have $\lambda_i\in\mathbb{R}$, $\forall i\in\mathcal{V}_n$.
	Then, second-order consensus is easily achieved for both protocols under any positive control gains.
	\begin{remark}
		The consensus criterion \eqref{eq:2.2.4} is not broadly known compared with \eqref{eq:2.2.5}.
		However, according to the proof of Lemma 2 in \cite{Zhu2009},
		it can be readily derived by using Hurwitz stability criteria to analyze the characteristic polynomial.
	\end{remark}
	Different from existing works,
	we aim to investigate whether the absolute velocity protocol \eqref{ab_pro} and the relative velocity protocol \eqref{re_pro} can achieve scalable second-order consensus for hierarchical group \eqref{eq:2.2.1}.
	Specifically, the scalability refers to two aspects.
	On the one hand, the scale of hierarchical group \eqref{eq:2.2.1} gradually grows.
	As more agents join,
	communication graphs of different sizes constitute a family of DAGs $\{\mathcal{G}_n:n\geq n_0\}$ called the \textit{hierarchical graph family},
	in which $n_0$ is the incipient group size, the network size $n$ is increasing and each DAG has a spanning tree and a linear extension ordering.
	On the other hand,
	as the group size grows,
	additional reverse edges constantly emerge.
	Then, we can obtain a new graph family $\{\bar{\mathcal{G}}_n:n\geq n_0\}$ called the \textit{mixed graph family},
	in which each graph $\bar{\mathcal{G}}_n$ is a mixed graph consisting of original DAG $\mathcal{G}_n$ with additional reverse edges,
	%generated by adding some reverse edges to the original DAG $\mathcal{G}_n$ in $\{\mathcal{G}_n:n\geq n_0\}$,
	and the associated Laplacian matrix $\bar{\mathcal{L}}_n$ takes the form in \eqref{L_bar}.
	For brevity, the hierarchical graph family and the mixed graph family are denoted by $\{\mathcal{G}_n\}$ and $\{\bar{\mathcal{G}}_n\}$, respectively.
	It should be stressed that control gains $\alpha$ and $\beta$ of all agents are  preset and cannot be re-tuned throughout $\{\bar{\mathcal{G}}_n\}$, regardless of whether the absolute velocity protocol \eqref{ab_pro} or the relative velocity protocol \eqref{re_pro} is employed.
	In addition, since hierarchical relations and topological structures reconstruct as the network size increases,
	$\bar{\mathcal{G}}_{n-1}$ needs not be a subgraph of $\bar{\mathcal{G}}_{n}$.
	%In brief, MAS \eqref{eq:2.2.1} deploys and extends in terms of $\{\bar{\mathcal{G}}_n\}$.
	
	With the above configurations and Assumption \ref{assumption:2},
	the scalable second-order consensus is defined as follows.
	\begin{definition}[Scalable second-order consensus]\label{def:3}
		For a given protocol,
		hierarchical group \eqref{eq:2.2.1} is said to achieve scalable second-order consensus on a mixed graph family $\{\bar{\mathcal{G}}_n\}$  if there exist fixed control gains $\alpha$ and $\beta$ such that hierarchical group \eqref{eq:2.2.1} achieves second-order consensus on any graph in $\{\bar{\mathcal{G}}_n\}$.
		In particular, if hierarchical group \eqref{eq:2.2.1} achieves scalable second-order consensus on any mixed graph family $\{\bar{\mathcal{G}}_n\}$,
		it is called completely scalable second-order consensus.
	\end{definition}
	\begin{remark}
		Compared with $\mathcal{G}_n$,
		the mixed graph $\bar{\mathcal{G}}_n$ may be no longer a DAG,
		which brings complex Laplacian eigenvalues $\bar{\lambda}_i$.
		In this case, second-order consensus could fail without adjusting control gains.
		However,
		due to the difference between consensus conditions \eqref{eq:2.2.4} and \eqref{eq:2.2.5} resulting from different underlying structures of protocols,
		the influence of reverse edges on consensus may exhibit different trends as the network size increases.
		Therefore, we wonder whether the consensus can be preserved under constantly changing network structures together with reverse edges by selecting the underlying structures of protocols rather than re-tuning the control gains.
	\end{remark}

	%%%%%%%%%%%%%%%%%%%%%%%%%%%%%%%%%%%%%%%%%%%%%%%%%%%%%%%%%%%%%%%
	\section{Main results}\label{Sec:4}
	%%%%%%%%%%%%%%%%%%%%%%%%%%%%%%%%%%%%%%%%%%%%%%%%%%%%%%%%%%%%%%%
	In this section, we analyze the scalable second-order consensus of the hierarchical group \eqref{eq:2.2.1} with the absolute velocity protocol \eqref{ab_pro} and the relative velocity protocol \eqref{re_pro}, respectively.
	Then, we propose a well-scalable hierarchical network for both protocols.

	The following theorem is given for the hierarchical group \eqref{eq:2.2.1} with the absolute velocity protocol \eqref{ab_pro}.
	\begin{theorem}\label{theorem1}
		Hierarchical group \eqref{eq:2.2.1} can achieve completely scalable second-order consensus with the absolute velocity protocol \eqref{ab_pro} under Assumption \ref{assumption:2}.
		%	For any $\{\bar{\mathcal{G}}_n\}$ subject to Assumptions \ref{assumption:2} and \ref{assumption:3},
		%	the MAS \eqref{eq:2.2.1}-\eqref{ab_pro} is scalably stable if $\frac{\beta^2}{\alpha}>2(\bar{a}q+\bar{b}\bar{m})$.
	\end{theorem}
	\begin{Proof}
		Note that each DAG $\mathcal{G}_n$ in the hierarchical graph family $\{\mathcal{G}_n\}$ has a spanning tree and is a subgraph of the corresponding mixed graph $\bar{\mathcal{G}}_n$.
		Thus, each mixed graph $\bar{\mathcal{G}}_n$ in the mixed graph family $\{\bar{\mathcal{G}}_n\}$ must has a spanning tree,
		whose Laplacian matrix $\bar{\mathcal{L}}_n$ has a simple zero eigenvalues and all the other non-zero eigenvalues have positive real parts \cite{Ren2005}.
		%    Thus, for any mixed graph $\bar{\mathcal{G}}_n$ in the mixed graph family $\{\bar{\mathcal{G}}_n\}$, 
		For convenience, let $\bar{\lambda}_i$ assume the form $0=\bar{\lambda}_1< \text{Re}(\bar{\lambda}_2)\leq\cdots\leq \text{Re}(\bar{\lambda}_n)$.
		%    the Laplacian matrix takes the form in \eqref{L_bar},
		%    and the eigenvalues of $\bar{\mathcal{L}}_n$ are denoted by $\bar{\lambda}_i$, $i\in\mathcal{V}_n$.
		%    Since the original DAG $\mathcal{G}_n$ is a subgraph of $\bar{\mathcal{G}}_n$ and has a spanning tree and ,
		%    $\bar{\mathcal{G}}_n$ must has a spanning tree.
		%    According to Lemma \ref{lem:1},
		%    $\bar{\lambda}_i$ assume the form $0=\bar{\lambda}_1< \text{Re}(\bar{\lambda}_2)\leq\cdots\leq \text{Re}(\bar{\lambda}_n)$.
		By inspecting \eqref{L_bar},
		we find that $\bar{l}_{ii}=\sum_{j\neq i}^n|\bar{l}_{ij}|=d_{ii}+p_{ii}$.
		Thereby, the Ger\v{s}gorin discs for rows of $\bar{\mathcal{L}}_n$ is
		$G(\bar{\mathcal{L}}_n)=\bigcup_{i=1}^n\{z\in\mathbb{C}:|z-(d_{ii}+p_{ii})|\leq d_{ii}+p_{ii}\}$,
		which are in the complex right half plane and tangent to the imaginary axis.
		According to Ger\v{s}gorin disc Theorem \cite{Horn2012},
		all eigenvalues $\bar{\lambda}_i$ of $\bar{\mathcal{L}}_n$ lie inside $G(\bar{\mathcal{L}}_n)$, i.e.,
		$\bar{\lambda}_i\in G(\bar{\mathcal{L}}_n)$, $\forall i\in \mathcal{V}_n$.
		
		For convenience, let $G_{\sigma,\delta}=\{z\in\mathbb{C}:|z-\sigma|\leq \delta, \sigma\in \mathbb{R}, \delta\in \mathbb{R}\}$ be a disc with radius $\delta$ and center $(\sigma,0)$ in the complex plane.
		%    Based on Assumptions \ref{assumption:2} and \ref{assumption:3},
		%    no matter how the original DAG $\mathcal{G}_n$ is designed and how the reverse edges are added,
		For any mixed graph $\bar{\mathcal{G}}_n$ subject to Assumption \ref{assumption:2},
		there hold $d_{ii}\leq d_{\max}\leq \zeta\bar{a}$ and $p_{ii}\leq \xi\bar{a}_r$ for any $i\in \mathcal{V}_n$,
		which give rise to $G(\bar{\mathcal{L}}_n)\subseteq G_{\mu,\mu}$,
		where $\mu=\zeta\bar{a}+\xi\bar{a}_r$.
		%    Thus, we have $\bar{\lambda}_i\in G_{\mu,\mu}$, $\forall i\in \mathcal{V}_n$.
		
		Let $\Omega=\{z\in\mathbb{C}:|z-\mu|=\mu\}$ be the boundary of the disc $G_{\mu,\mu}$.
		Since $\Omega$ is symmetric about the real axis,
		for any complex number $\omega\neq0$ in $G_{\mu,\mu}$,
		there exist two conjugate complex numbers $c_1,c_2\in\{z\in\Omega: \text{Re}(z)=\text{Re}(\omega)\}$ such that
		$$\frac{\text{Im}(c_1)^2}{\text{Re}(\omega)}=\frac{\text{Im}(c_2)^2}{\text{Re}(\omega)}\geq\frac{\text{Im}(\omega)^2}{\text{Re}(\omega)},$$
		which means that
		$$\arg\sup_{\omega\in G_{\mu,\mu}}\frac{\text{Im}(\omega)^2}{\text{Re}(\omega)}\in\Omega.$$
		Therefore, we only need to focus on the complex number $\omega\neq 0$ on the boundary $\Omega$. 
		
		If $\mu\leq \text{Re}(\omega)\leq 2\mu$, $\omega\in \Omega$,
		it follows that $\text{Im}(\omega)^2$ is decreasing with $\text{Re}(\omega)$.
		Then, we obtain 
		\begin{equation}\label{mu}
			\sup_{\substack{\omega\in \Omega\\ \mu\leq \text{Re}(\omega)\leq 2\mu}}\frac{\text{Im}(\omega)^2}{\text{Re}(\omega)}=\mu.
		\end{equation}
		If $0<\text{Re}(\omega)<\mu$, $\omega\in \Omega$,
		we draw two auxiliary lines in the complex plane, which are shown as the dash lines in Fig. \ref{fig1}.
		\begin{figure}[!ht]\centering
			\includegraphics[width=0.27\textwidth,height=0.24\textwidth]{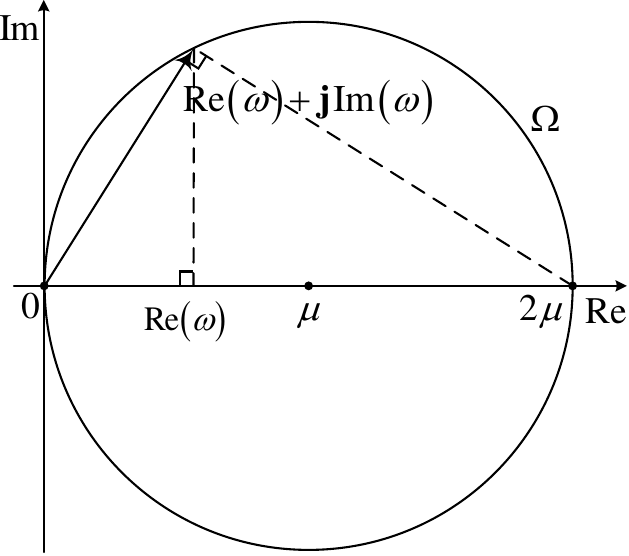}
			\caption{Diagram for the case $0<\text{Re}(\omega)<\mu$.\label{fig1}}
		\end{figure}
		One of the lines connects $(\text{Re}(\omega),\text{Im}(\omega))$ and $(2\mu,0)$. 
		The other one is perpendicular to the real axis and passes through $(\text{Re}(\omega),\text{Im}(\omega))$.
		Obviously, it is observed that $\frac{\text{Re}(\omega)}{|\omega|}=\frac{|\omega|}{2\mu}$.
		Thus, we can get
		$\frac{\text{Im}(\omega)^2}{\text{Re}(\omega)}=2\mu-\text{Re}(\omega)$
		which is decreasing with $\text{Re}(\omega)$.
		If follows that
		\begin{equation}\label{2mu}
			\sup_{\substack{\omega\in \Omega\\ 0<\text{Re}(\omega)<\mu}}\frac{\text{Im}(\omega)^2}{\text{Re}(\omega)}=2\mu.
		\end{equation}
		Combining
		$G(\bar{\mathcal{L}}_n)\subseteq G_{\mu,\mu}$,
		$\arg\sup_{\omega\in G_{\mu,\mu}}\frac{\text{Im}(\omega)^2}{\text{Re}(\omega)}\in\Omega$,
		\eqref{mu} and \eqref{2mu},
		we can obtain
		\begin{equation*}
			\max_{\bar{\lambda}_i\neq0}\frac{\text{Im}(\bar{\lambda}_i)^2}{\text{Re}(\bar{\lambda}_i)}\leq
			\sup_{\substack{\omega\neq 0 \\ \omega\in G_{\mu,\mu}}}\frac{\text{Im}(\omega)^2}{\text{Re}(\omega)}\leq
			\sup_{\substack{\omega\neq 0\\ \omega\in\Omega}}\frac{\text{Im}(\omega)^2}{\text{Re}(\omega)}=2\mu.
		\end{equation*}
		Therefore, according to \eqref{eq:2.2.4},
		if control gains are preset as $\frac{\beta^2}{\alpha}>2(\zeta\bar{a}+\xi\bar{a}_r)$,
		hierarchical group \eqref{eq:2.2.1}-\eqref{ab_pro} reaches second-order consensus for any mixed graph $\bar{\mathcal{G}}_n$ in any mixed graph family $\{\bar{\mathcal{G}}_n\}$.
		That is, hierarchical group \eqref{eq:2.2.1} can achieve completely scalable second-order consensus with the absolute velocity protocol \eqref{ab_pro}.
	\end{Proof}
	%\begin{remark}
	%	Although the completely scalable second-order consensus is defined for the mixed graph families,
	%	Theorem \ref{theorem1} also applies to the general directed graph families.
	%	%	Consider a family of weighted directed graph $\{\mathcal{G}_n\}$ with finite maximum degree, i.e., $d_{\max}\leq \bar{d}$.
	%	Similar to the proof of Theorem \ref{theorem1},
	%	we can verify that
	%	for any weighted directed graph family $\{\mathcal{G}_n\}$ with finite maximum degree, i.e., $d_{\max}\leq \bar{d}$,
	%	system \eqref{eq:2.2.1}-\eqref{ab_pro} achieves second-order consensus on any weighted directed graph in $\{\mathcal{G}_n\}$ if control gains are preset as $\frac{\beta^2}{\alpha}>2\bar{d}$.
	%	Therefore, completely scalable second-order consensus is still obtained. 
	%\end{remark}
	By contrast, in the following theorem,
	we prove that the relative velocity protocol \eqref{re_pro} fails to reach the completely scalable second-order consensus.
	
	\begin{theorem}\label{theorem2}
		Hierarchical group \eqref{eq:2.2.1} cannot achieve completely scalable second-order consensus with the relative velocity protocol \eqref{re_pro} under Assumption \ref{assumption:2}.
	\end{theorem}
	\begin{Proof}
		According to Definition \ref{def:3},
		we just need to prove that there exists a class of mixed graph family such that hierarchical group \eqref{eq:2.2.1}-\eqref{re_pro} cannot achieve scalable second-order consensus on it.
		
		Consider a family of hierarchical graphs $\{P_n\}$,
		in which each $P_n$ is a directed path graph with a linear extension ordering and unit edge weights.
		In particular, the linear extension ordering of $P_n$ is unique,
		which is the same as the order that the agents appear along the path.
		Suppose that there is only single reverse edge taking unit weight in each $P_n$, which points from agent $m$ to agent $q$ ($1=q<m\leq n$),
		and the span of the reverse edge $s=m-q$ is increasing with the network size $n$.
		Then, $P_n$ and the above extra reverse edge constitute a mixed graph  $\bar{P}_n$.
		The mixed graph family is represented as $\{\bar{P}_n\}$.
		
		For each $\bar{P}_n$ in $\{\bar{P}_n\}$, 
		the reverse edge always includes the root node 1.
		Thus, the Laplacian matrix of $\bar{P}_n$ in the form of \eqref{L_bar}
		can be written as
		$
		\bar{\mathcal{L}}_n=\begin{bmatrix}
			\bar{\Theta} & \\
			* & L_2\end{bmatrix},
		$
		where $L_2\in\mathbb{R}^{(n-m)\times(n-m)}$ is a lower triangular matrix whose eigenvalues are all real,
		and $\bar{\Theta}=\Theta+\Delta\in\mathbb{R}^{(s+1)\times(s+1)}$ is exactly the Laplacian matrix of a directed ring graph with $s+1$ vertices and unit edge weights,
		whose non-zero eigenvalues are $\gamma_i=1-\phi_i+\mathbf{j}\psi_i$, $i=1,\dots,s$
		\cite{Agaev2010},
		where $\phi_i=\cos(\frac{2\pi i}{s+1})$ and $\psi_i=\sin(\frac{2\pi i}{s+1})$.
		A simple calculation yields
		\begin{equation}
			\begin{aligned}
				\frac{\text{Im}(\gamma_i)^2}{\text{Re}(\gamma_i)|\gamma_i|^2}
				&=\frac{\psi_i^2}{[1-\phi_i]\cdot[(1-\phi_i)^2+\psi_i^2]}\\
				&=\frac{1+\phi_i}{2\cdot[1-\phi_i]}=\frac{1}{2}\cot^2(\frac{\pi i}{s+1}).
			\end{aligned}
		\end{equation}
		It can be readily concluded that
		\begin{equation*}
			\max_{\bar{\lambda}_i\neq 0}\frac{\text{Im}(\bar{\lambda}_i)^2}{\text{Re}(\bar{\lambda}_i)|\bar{\lambda}_i|^2}
			=\max_{i=1,\dots,s}\frac{\text{Im}(\gamma_i)^2}{\text{Re}(\gamma_i)|\gamma_i|^2}
			=\frac{1}{2}\cot^2(\frac{\pi}{s+1}).
		\end{equation*}
		Note that the span of the reverse edge $s$ is increasing with the network size $n$.
		%	$2)$ $h>1$
		%	In this case, the reverse edge excludes the root node 1.
		Therefore, combining with \eqref{eq:2.2.5},
		there are no fixed control gains to preserve consensus as the network size grows.
		According to Definition \ref{def:3},
		hierarchical group \eqref{eq:2.2.1}-\eqref{re_pro} cannot reach scalable second-order consensus on $\{\bar{P}_n\}$.
		Thereby, completely scalable second-order consensus is naturally failed.
	\end{Proof}
	\begin{remark}
		In fact, provided that the single reverse edge considered in the proof of Theorem \ref{theorem2} excludes the root node, i.e., $1<q<m\leq n$,
		hierarchical group \eqref{eq:2.2.1}-\eqref{re_pro} still cannot achieve scalable second-order consensus on $\{\bar{P}_n\}$.
		In conclusion,
		as long as the span of the reverse edge is increasing with the network size,
		consensus could fail.
		This conclusion reveals the serious consequence that an individual bypasses its immediate leader and reports to a higher-level one.
	\end{remark}

	We are surprised that if the hierarchical group \eqref{eq:2.2.1}-\eqref{re_pro} deploys and expands in terms of a path structure,
	one reverse edge is enough to break the consensus.
	This is in sharp contrast with the hierarchical group \eqref{eq:2.2.1} taking the absolute velocity protocol \eqref{re_pro}.
	As shown in the proof of Theorem \ref{theorem1},
	for the absolute velocity protocol \eqref{ab_pro},
	there always exist fixed control gains such that the hierarchical group \eqref{eq:2.2.1} reaches consensus no matter how the topological structures and the reverse edges change.
	Our main results Theorem \ref{theorem1} and \ref{theorem2} reveal that the  absolute velocity protocol \eqref{ab_pro} outperforms the relative velocity protocol \eqref{re_pro} in terms of scalability of second-order consensus for hierarchical groups.
	In addition, this finding also theoretically tells that the local interaction rules among members affect the scalability of coordination behavior.
	
	Besides $\{\bar{P}_n\}$,
	there certainly exist other mixed graph families such that the relative velocity protocol \eqref{re_pro} fails to attain scalable second-order consensus.
	However, we are more interested in designing a common scalable hierarchical structure for both the absolute velocity protocol \eqref{ab_pro} and the relative velocity protocol\eqref{re_pro}.
	
	Consider a hierarchical graph family $\{S_n\}$ consisting of a sequence of directed star graphs with increasing network size $n$.
	Each $S_n$ has a linear extension ordering and $n$ nodes including a hub node and $n-1$ fringe nodes.
	Every fringe node admits a directed edge with the same weight $\rho$ pointing from the hub node to it.
	Here, 
	the weights and the number of reverse edges are with no restrictions.
	Graph family $\{S_n\}$ with additional reverse edges is represented as the mixed graph family $\{\bar{S}_n\}$.
	Then, we obtain the following result.
	\begin{theorem}
		Hierarchical group \eqref{eq:2.2.1} can achieve scalable second-order consensus on $\{\bar{S}_n\}$ with either the absolute velocity protocol \eqref{ab_pro} or the relative velocity protocol \eqref{re_pro}.
	\end{theorem}
	\begin{Proof}
		For each $\bar{S}_n$ in $\{\bar{S}_n\}$,
		the characteristic polynomial of the Laplacian matrix $\bar{\mathcal{L}}_n$ in the form \eqref{L_bar} can be computed as
		\begin{equation*}
				|\lambda I_n-\bar{\mathcal{L}}_n|
				=\begin{vmatrix}
					\lambda-p_{11} & \bar{a}_{12} & \bar{a}_{13} & \dots & \bar{a}_{1n}\\
					\rho & \lambda-(\rho+p_{22}) & \bar{a}_{23} & \dots & \bar{a}_{2n}\\
					\rho & 0 & \lambda-(\rho+p_{33}) & \dots & \bar{a}_{3n}\\
					\vdots & \vdots & \vdots & \ddots & \vdots\\
					\rho & 0 & 0 & \dots & \lambda-\rho
				\end{vmatrix}
				\triangleq|\hat{\mathcal{L}}_0(\lambda)|,
		\end{equation*}
		where $\bar{a}_{ij}\geq 0$ ($i<j$) and $p_{ii}=\sum_{j=i+1}^{n}\bar{a}_{ij}$, $i=1,\dots,n-1$.
		Recall that $\bar{a}_{ij}>0$ implies a reverse edge with weight $\bar{a}_{ij}$ from $j$ to $i$,
		and $\bar{a}_{ij}=0$ means no such reverse edge.
		
		Then, we make a series of elementary row and column operations for $\hat{\mathcal{L}}_0(\lambda)$.
		Firstly, starting with $j=1$,
		adding each column after the $j_\mathrm{th}$ column of $\hat{\mathcal{L}}_{j-1}(\lambda)$ to the $j_\mathrm{th}$ column yields a new matrix $\hat{\mathcal{L}}_j(\lambda)$.
		Subsequently,
		repeating the above elementary column operations for $j=2,\dots,n-1$ in turn,
		we can obtain a series of matrices $\hat{\mathcal{L}}_0(\lambda),\dots,\hat{\mathcal{L}}_{n-1}(\lambda)$,
		where
		\begin{equation*}
			\hat{\mathcal{L}}_{n-1}(\lambda)
			=\begin{bmatrix}
				\lambda & p_{11} & \ast &  \dots & \ast & \bar{a}_{1n}\\
				\lambda & \lambda-\rho & p_{22} & \dots & \ast & \bar{a}_{2n}\\
				\lambda & \lambda-\rho & \lambda-\rho & \dots & \ast & \bar{a}_{3n}\\
				%			\lambda & \lambda-\rho & \lambda-\rho & \lambda-\rho & \dots & \ast & \bar{a}_{4n}\\
				\vdots & \vdots & \vdots & \ddots & \vdots & \vdots\\
				\lambda & \lambda-\rho & \lambda-\rho & \dots & \lambda-\rho & p_{(n-1)(n-1)}\\
				\lambda & \lambda-\rho &  \lambda-\rho  & \dots & \lambda-\rho & \lambda-\rho
			\end{bmatrix}.
		\end{equation*}

		Finally, every row of $\hat{\mathcal{L}}_{n-1}(\lambda)$ except the last row minus the last row which gives rise to the matrix $\tilde{\mathcal{L}}(\lambda)$.
		Therefore, we obtain
		\begin{equation*}
				|\hat{\mathcal{L}}_0(\lambda)|=|\hat{\mathcal{L}}_{n-1}(\lambda)|=|\tilde{\mathcal{L}}(\lambda)|
				=\begin{vmatrix}
					0 & \rho+p_{11}-\lambda & * & \dots & *\\
					0 & 0 & \rho+p_{22}-\lambda & \dots & *\\
					\vdots & \vdots & \vdots & \ddots & \vdots\\
					0 & 0 & 0 & \dots & \rho+p_{(n-1)(n-1)}-\lambda\\
					\lambda & \lambda-\rho & \lambda-\rho & \dots & \lambda-\rho
				\end{vmatrix}
				=\lambda\sum_{i=1}^{n-1}[\lambda-(\rho+p_{ii})].
		\end{equation*}
		
		It is obvious that all non-zero Laplacian eigenvalues are real for networks of any sizes with reverse edges of any number and any weight.
		Thus, according to \eqref{eq:2.2.4} and \eqref{eq:2.2.5},
		for any positive preset control gains,
		hierarchical group \eqref{eq:2.2.1} is able to reach consensus with either the protocol \eqref{ab_pro} or protocol \eqref{re_pro},
		that is,
		scalable second-order consensus can be achieved on $\{\bar{S}_n\}$ for either of the protocols.
	\end{Proof}
	\begin{remark}
		Compared with the directed path graph,
		the directed star graph exhibits remarkable scalability under multiple reverse edges.
		This appears to imply a relation between the scalability and the height of the network which refers to the length of a longest path from the root node \citep{Gross2003}.
		But it needs to be further verified by graph-theoretic analysis.
	\end{remark}
	
	%%%%%%%%%%%%%%%%%%%%%%%%%%%%%%%%%%%%%%%%%%%%%%%%%%%%%%%%%
	\section{Conclusion}\label{Sec:5}
	%%%%%%%%%%%%%%%%%%%%%%%%%%%%%%%%%%%%%%%%%%%%%%%%%%%%%%%%%
	
	This study explored the scalable second-order consensus of hierarchical groups composed of a group of agents with double-integrator dynamics on DAGs.
	More specifically, as the network size increases and in the presence of reverse edges,
	we investigated which one of the absolute and the relative velocity protocols can preserve the system consensus property without tuning the control gains.
	It was proved that the absolute velocity protocol can reach completely scalable second-order consensus but the relative velocity protocol cannot.
	The scalable second-order consensus for the relative velocity protocol only happens on certain mixed graph families.
	This finding seems to explain the phenomenon that the local interaction rules are changed for larger scale of coordination behaviors \cite{Detrain2006}.
	Furthermore, we proposed a hierarchical architecture, i.e., the directed star graphs, on which second-order consensus can be reached for networks of any sizes with reverse edges of any number and any weight.
	We conjectured that the outstanding scalability of the structure may be related to height of the networks,
	which deserves to be further verified in future works.

	%%%%%%%%%%%%%%%%%%%%%%%%%%%%%%%%%%%%%%%%%%%%%%%%%%%%%%%%%%%%%%%%%%%%%%

	
	
	
	
	%\begin{thebibliography}{00}
	%%% Text of bibliographic item
	%
	%\bibitem{}
	%
	%\end{thebibliography}
	
	\bibliography{references}

\begin{thebibliography}{10}
\expandafter\ifx\csname url\endcsname\relax
  \def\url#1{\texttt{#1}}\fi
\expandafter\ifx\csname urlprefix\endcsname\relax\def\urlprefix{URL }\fi
\expandafter\ifx\csname href\endcsname\relax
  \def\href#1#2{#2} \def\path#1{#1}\fi

\bibitem{Nagy2010}
M.~Nagy, Z.~{\'A}kos, D.~Biro, T.~Vicsek, Hierarchical group dynamics in pigeon
  flocks, Nature 464 (2010) 890--893.

\bibitem{Franks1983}
N.~R. Franks, E.~Scovell, Dominance and reproductive success among slave-making
  worker ants, Nature 304 (1983) 724--725.

\bibitem{Sapolsky2005}
R.~M. Sapolsky, The influence of social hierarchy on primate health, Science
  308~(5722) (2005) 648--652.

\bibitem{Shimoji2014}
H.~Shimoji, M.~S. Abe, K.~Tsuji, N.~Masuda, Global network structure of
  dominance hierarchy of ant workers, Journal of the Royal Society Interface
  11~(99) (2014) 20140599.

\bibitem{Gross2003}
J.~L. Gross, P.~Zhang, J.~Yellen, Handbook of graph theory, CRC press, 2003.

\bibitem{Shao2016}
J.~Shao, J.~Qin, A.~N. Bishop, T.-Z. Huang, W.~X. Zheng, A novel analysis on
  the efficiency of hierarchy among leader-following systems, Automatica 73
  (2016) 215--222.

\bibitem{Shao2018}
J.~Shao, W.~X. Zheng, T.-Z. Huang, A.~N. Bishop, On leader–follower consensus
  with switching topologies: An analysis inspired by pigeon hierarchies, IEEE
  Transactions on Automatic Control 63~(10) (2018) 3588--3593.

\bibitem{Detrain2006}
C.~Detrain, J.-L. Deneubourg, Self-organized structures in a superorganism: do
  ants ``behave" like molecules?, Physics of Life Reviews 3~(3) (2006)
  162--187.

\bibitem{Douglas2017}
P.~H. Douglas, A.-C.~N. Ngomo, G.~Hohmann, A novel approach for dominance
  assessment in gregarious species: Adagio, Animal Behaviour 123 (2017) 21--32.

\bibitem{Zhang2017}
H.-T. Zhang, Z.~Chen, X.~Mo, Effect of adding edges to consensus networks with
  directed acyclic graphs, IEEE Transactions on Automatic Control 62~(9) (2017)
  4891--4897.

\bibitem{Mo2019}
X.~Mo, Z.~Chen, H.-T. Zhang, Effects of adding a reverse edge across a stem in
  a directed acyclic graph, Automatica 103 (2019) 254--260.

\bibitem{Hao2019}
Y.~Hao, Q.~Wang, Z.~Duan, G.~Chen, The role of reverse edges on consensus
  performance of chain networks, IEEE Transactions on Systems, Man, and
  Cybernetics: Systems 51~(3) (2019) 1757--1765.

\bibitem{Zhang2022}
H.-T. Zhang, H.~Cao, Z.~Chen, A necessary and sufficient condition of an
  interfering reverse edge for a directed acyclic graph, IEEE Transactions on
  Automatic Control 67~(9) (2022) 4885--4891.

\bibitem{Zhu2009}
J.~Zhu, Y.-P. Tian, J.~Kuang, On the general consensus protocol of multi-agent
  systems with double-integrator dynamics, Linear Algebra and its Applications
  431~(5-7) (2009) 701--715.

\bibitem{Yu2010}
W.~Yu, G.~Chen, M.~Cao, Some necessary and sufficient conditions for
  second-order consensus in multi-agent dynamical systems, Automatica 46~(6)
  (2010) 1089--1095.

\bibitem{Dubey2022}
A.~K. Dubey, D.~Mukherjee, K.~Arya, Consensus of double integrators over a
  chain with reverse edges, IFAC-PapersOnLine 55~(13) (2022) 19--24.

\bibitem{Dubey2023}
A.~K. \vspace{0mm}Dubey, D.~Mukherjee, K.~Arya, Consensus of double integrators
  in presence of a reverse edge in a chain: Analysis and design, in:
  Proceedings of 2023 American Control Conference, 2023, pp. 411--416.

\bibitem{Fischhoff2007}
I.~R. Fischhoff, S.~R. Sundaresan, J.~Cordingley, H.~M. Larkin, M.-J. Sellier,
  D.~I. Rubenstein, Social relationships and reproductive state influence
  leadership roles in movements of plains zebra, equus burchellii, Animal
  Behaviour 73~(5) (2007) 825--831.

\bibitem{Tegling2023}
E.~Tegling, B.~Bamieh, H.~Sandberg, Scale fragilities in localized consensus
  dynamics, Automatica 153 (2023) 111046.

\bibitem{Ren2007}
W.~Ren, E.~Atkins, Distributed multi-vehicle coordinated control via local
  information exchange, International Journal of Robust and Nonlinear Control
  17~(10-11) (2007) 1002--1033.

\bibitem{Xie2007}
G.~Xie, L.~Wang, Consensus control for a class of networks of dynamic agents,
  International Journal of Robust and Nonlinear Control 17~(10-11) (2007)
  941--959.

\bibitem{Zheng2011}
Y.~Zheng, Y.~Zhu, L.~Wang, Consensus of heterogeneous multi-agent systems, IET
  Control Theory \& Applications 5~(16) (2011) 1881--1888.

\bibitem{Li2015}
H.~Li, X.~Liao, T.~Huang, W.~Zhu, Event-triggering sampling based
  leader-following consensus in second-order multi-agent systems, IEEE
  Transactions on Automatic Control 60~(7) (2015) 1998--2003.

\bibitem{Hou2017}
W.~Hou, M.~Fu, H.~Zhang, Z.~Wu, Consensus conditions for general second-order
  multi-agent systems with communication delay, Automatica 75 (2017) 293--298.

\bibitem{Zheng2019}
Y.~Zheng, Q.~Zhao, J.~Ma, L.~Wang, Second-order consensus of hybrid multi-agent
  systems, Systems \& Control Letters 125 (2019) 51--58.

\bibitem{Zhao2020-1}
Q.~Zhao, Y.~Zheng, Y.~Zhu, Consensus of hybrid multi-agent systems with
  heterogeneous dynamics, International Journal of Control 93~(12) (2020)
  2848--2858.

\bibitem{Xu2022}
W.~Xu, J.~Kurths, G.~Wen, X.~Yu, Resilient event-triggered control strategies
  for second-order consensus, IEEE Transactions on Automatic Control 67~(8)
  (2022) 4226--4233.

\bibitem{Zhou2023}
L.~Zhou, J.~Liu, Y.~Zheng, F.~Xiao, J.~Xi, Game-based consensus of hybrid
  multiagent systems, IEEE Transactions on Cybernetics 53~(8) (2023)
  5346--5357.

\bibitem{Wang2022}
J.~Wang, L.~Zhou, D.~Zhang, J.~Liu, Y.~Zheng, Protocol selection for
  second-order consensus against disturbance, arXiv preprint: 2212.05240
  (2022).

\bibitem{Vicsek1995}
T.~Vicsek, A.~Czir{\'o}k, E.~Ben-Jacob, I.~Cohen, O.~Shochet, Novel type of
  phase transition in a system of self-driven particles, Physical Review
  Letters 75~(6) (1995) 1226--1229.

\bibitem{Mora2016}
T.~Mora, A.~M. Walczak, L.~Del~Castello, F.~Ginelli, S.~Melillo, L.~Parisi,
  M.~Viale, A.~Cavagna, I.~Giardina, Local equilibrium in bird flocks, Nature
  Physics 12 (2016) 1153--1157.

\bibitem{Ren2005}
W.~Ren, R.~W. Beard, Consensus seeking in multiagent systems under dynamically
  changing interaction topologies, IEEE Transactions on Automatic Control
  50~(5) (2005) 655--661.

\bibitem{Horn2012}
R.~A. Horn, C.~R. Johnson, Matrix analysis, Cambridge university press, 2012.

\bibitem{Agaev2010}
R.~Agaev, P.~Chebotarev, Which digraphs with ring structure are essentially
  cyclic?, Advances in Applied Mathematics 45~(2) (2010) 232--251.

\end{thebibliography}
\end{document}